\documentclass[10pt,journal]{IEEEtran}
\usepackage{cite}

\usepackage[utf8]{inputenc} 
\usepackage[T1]{fontenc}
\usepackage[euler]{textgreek}

\usepackage{enumitem}

\usepackage{microtype}  
\usepackage{booktabs} 

\usepackage[top=2cm, bottom=2cm, left=3cm, right=3cm]{geometry}

\usepackage{url} % Pour ecrire des adresses cliquables.
\usepackage{lmodern} % Pour changer le pack de police.

\usepackage{graphicx,dblfloatfix}
\usepackage{float}
\usepackage{subcaption}

\usepackage{wrapfig} % wrapper une figure entre du text
\usepackage{amsmath}

\usepackage{algorithm}
\usepackage{algpseudocode}

\usepackage{epstopdf}

\begin{document}

\title{Cooperative Evaluation of the Cause of Urban Traffic Congestion via Connected Vehicles }

\author{Ranwa~Al Mallah, Alejandro~Quintero, and~Bilal~Farooq% <-this % stops a space
\thanks{R. Al Mallah and A. Quintero are with the Department of Computer Science, École Polytechnique de Montréal, CANADA, e-mails: ranwa.al-mallah@polymtl.ca and alejandro.quintero@polymtl.ca.}% <-this % stops a space
\thanks{B. Farooq is with the Laboratory of Innovations in Transportation, Ryerson University, Toronto, CANADA, email: bilal.farooq@ryerson.ca.}}

\maketitle

\begin{abstract}
We developed a distributed data mining system to elaborate a decision concerning the cause of urban traffic congestion via emerging connected vehicle (CV) technology. We observe this complex phenomena through the interactions between vehicles exchanging messages via Vehicle to Vehicle (V2V) communication. Results are based on real-time simulation generated scenarios extended from the real-world traffic Travel and Activity PAtterns Simulation (TAPAS) Cologne scenario. We evaluate a Voting Procedure (VP) useful for obtaining deeper insights using cooperation between vehicles, Belief Functions (BF) aim at improving representation of information and a Data Association Technique (DAT) aiming at data mining and extracting the association rules from the messages exchanged. Methods are tested and compared using a microscopic urban mobility simulator, SUMO and a network simulator, ns-2, for the simulation of communication between CVs. Compared to the Back-Propagation algorithm (BP) extensively used in the past literature, our performance evaluation shows that the proposed methods enhance the estimation of the cause of congestion by 48\% for the proposed VP, 58\% for the BF, 71\% for the DAT and 70\% for \textbeta-DAT. The methods also enhance detection time from 7.09\% to 10.3\%, and \textbeta-DAT outperforms BP by approximately 1.25\% less false alarms triggered by the network, which can be significant in the context of real-time decision making. We show that a market penetration rate between 63\% and 75\% is enough to ensure satisfactory performance. \newline
\end{abstract}


\section{INTRODUCTION} 
\IEEEPARstart{W}{ith} the increasing number of vehicles and limited road network expansion, the urban traffic congestion is growing at an alarming rate. Due to the high complexity and uncertainty of contemporary transportation systems, traditional traffic data collection and estimation tools fail to capture the detailed dynamics in real-time. With the progress in information and communication technologies,  Vehicular Ad hoc NETworks (VANET) known as Connected Vehicles (CVs) aims at building an intelligent public transportation system based on real-time information \cite{lu2014connected, hu2017cyber}. 

A large body of work has already focused on the congestion detection problem using CVs \cite{6857980}. Recently, a framework was proposed to further characterize the congestion detected. Chow et al. \cite{chow2014empirical} classified the congestion classification problem into its components in urban traffic. Al Mallah et al. \cite{7837642} proposed a model where if a vehicle detects congestion, it is able to estimate its causes based on macroscopic and microscopic traffic variables the vehicle collected along its trajectory. 

Several unresolved problems exist for CVs-based congestion classification. Firstly, each vehicle classifies individually the cause of congestion based on data from vehicles it encountered along its route. If one vehicle sends a false alarms, it spreads uncertainty among vehicles. The side effects of false alarms on the congestion level are a serious challenge because sending false information disrupts the proper network operation. This behaviour is a threat to the traffic network because in terms of security, it is comparable to an attacker injecting false information. This makes exploring the cause of congestion at a vehicle level a limited solution. Honest vehicles may behave as malicious users without their intent but because of the scheme. 

Secondly, in the schemes proposed in the literature each vehicle classifies individually the type of event based on its personal information \cite{BOUFARAH2016299}. It has been well established that the vehicular traffic is multifaceted, a vehicle has only partial knowledge about the road condition, and a vehicle knows only to some degree the traffic condition surrounding it. This decreases the estimation accuracy of the cause of congestion. Schemes should be implemented to obtain deeper insight on the cause of traffic congestion because inaccurately estimating the cause of congestion misleads other vehicles as well as traffic controllers and leads to devastating consequences. It has similar consequences on the network as false alarms. 
 
Thirdly, vehicles are repeatedly faced with situations where they encounter congestion and have to repeatedly determine its cause based on the variables they collected. They should be intelligent enough to learn from their experiences. Every situation should be a suite of instances learned for better decision making because the monitoring currently done by the proposed schemes does not allow for summarizing valuable knowledge. For a given situation, the classification algorithm proposed in the literature returns the cause with the greatest probability as the most likely cause of urban traffic congestion \cite{7837642}. A classification algorithm is a procedure for selecting a hypothesis from a set of alternatives that best fits a set of observation: a classifier returns only the cause with the highest probability. We propose that we can make use of the other probabilities computed by the classifier to extract more knowledge. 

The goal is to obtain deeper insights on the cause of traffic congestion using cooperation between CVs. Individual vehicle assessment is not enough, the next step is to elaborate a decision collectively. In the current state, if an event received by a vehicle is a false alarm, the algorithm will fuse the information with others located on a same road segment. An evaluation process has to take place before data fusion. We add this layer to address the vulnerability of fusion algorithms and to lower the side effects of false alarms because the approaches proposed in the literature fail to process the data before fusion. The methods we implement consist of a voting procedure (VP) and belief functions (BF). Also, we propose that each vehicle represents its uncertainty about the cause of congestion in a vector of probabilities associated to each of the possible causes of congestion before exchanging the vector with CVs. We then explore the collected vectors for learning purposes by building a dataset and extracting relationships via a data mining technique (DAT) to develop useful patterns. Particularly, this data analysis builds models capable of machine learning.

We consider a realistic map configuration of the city of Cologne in the evaluation of our methods. Compared to the Back-Propagation algorithm (BP) proposed in the literature \cite{7837642}, the proposed techniques enhance the estimation accuracy by 48\% for the VP, 58\% for the BF, 71\% for the DAT, and 70\% for the adapted data association technique (\textbeta-DAT). The methods also enhance the detection time by 10.3\% for VP, 9.40\% for BF, 9.45 for DAT and 7.09\% for \textbeta-DAT. \textbeta-DAT outperforms BP by approximately 1.25\% less false alarms triggered by the network. The methods also require only 63\% penetration rate to obtain the full benefits of V2V. Knowing the root causes of congestion that are affecting their facilities will enable road authorities to make more informed decisions about how to best reroute traffic, change lane priorities and modify traffic light sequences.

The contributions of this paper are summarized as follows:

\begin{itemize}

\item Addition of a cooperation process and an evaluation layer before fusion can take place in order to lower false alarms comparable to security threats on the traffic network;

\item Generation of a dataset for association rules mining to extract more knowledge with CVs being the communication architecture for decentralized exchange of information between the cooperating vehicles;

\item Validation of the methods using a microscopic urban mobility simulator, SUMO \cite{krajzewicz2002sumo} and a network simulator, ns-2 \cite{rehmaninetwork}, for the simulation of communication between CVs and evaluation of the necessary market penetration rate of CVs on the performance of the methods for transportation to obtain the full benefits of Vehicle-to-Vehicle (V2V) communications.

\end{itemize}

To accurately and systematically assess the value of the proposed methods, we simulate experiments on the real-world traffic Travel and Activity PAtterns Simulation (TAPAS) Cologne scenario \cite{uppoor2011large}. A review of related studies is provided in Section II. The methods are detailed in Section III. In Section IV, we provide results, analysis and discussion. Finally, conclusion and future work are outlined in Section V.

\section{RELATED WORK}

A Traffic Management System (TMS) consists of a set of complementary phases, each of which plays a specific role in ensuring efficient monitoring and control of the traffic flow in the city \cite{6857980}. In the first phase, traffic data can be collected from fixed monitoring equipment, but discovering the dynamic properties of the traffic is a difficult task due to the sparseness of the deployed equipments. New technologies can be used to improve the accuracy, timeliness, and cost efficiency of data collection \cite{6857980}. Connected vehicles of the VANET are scalable enough to enable better control of the traffic flow and enhance management of large cities’ road networks. 

In the second phase of a TMS, data fusion algorithms take the data collected from the information dissemination process and aggregate it to extract useful traffic information. Particularly, a large portion of literature has been proposed for the distributed data fusion for uncertain reasoning in ad hoc and dynamical networks \cite{4632299} \cite{BOUFARAH2016299}. In \cite{4632299}, they introduced belief functions to combine and fuse data in vehicle for the management of uncertainties about events in vehicular networks. Belief regarding the presence of an event on a geographical point is obtained by: discounting neighbouring information according to their distance from the point. The authors propose to use the cautious combination rule to fuse information located on the same road segment. In \cite{BOUFARAH2016299}, strategies to fuse acquired information consider message aging of local events. There are two major drawbacks to these approaches. Firstly, if the event received by the fusion algorithm is a false alarm, the algorithm will fuse the obtained information with others located on a same road segment and spread uncertainty among vehicles and this in turn might cause more congestion. Secondly, the approaches fail to process the data before fusion. The methods manage uncertainties by combining degrees of confidence about events reported in exchanged messages based on attributes such as the geographic distance of the event from the receiver and message aging of the event. For validation, Cherfaoui et al. \cite{4632299} used a Matlab simulator where roads are divided into segments and Bou Farah et al. \cite{BOUFARAH2016299} suggested that coupling their method with an ad hoc network simulator will be an added value for validation. 

In our study, we use CVs to monitor the traffic parameters experienced by a vehicle along its trajectory to enable granular and timely monitoring of events that occur on the roads. Our work concerns V2V communication mode where vehicles do not use any centralized unit to build their own information assembly. Unlike the approaches proposed in the literature, we deal with uncertainties via cooperation between vehicles on the road segment before data fusion can take place. Moreover, data mining techniques such as clustering, association, classification, have been applied in VANET to extract useful patterns and information \cite{8258162}. Particularly, we explore collected data between CVs for extracting relationships via data mining techniques. 
 
\section{DATA MINING METHODS}

Each vehicle is equipped with a method to detect excessive congestion and a classification algorithm able to attribute a possible cause to it, as in \cite{7837642}. We then propose that each vehicle represent its uncertainty about the cause of congestion in a vector of probabilities associated to each of the possible causes of:

C = [P$\tiny incident$, P$\tiny workzone$, P$\tiny weather$, P$\tiny specialevent$, P$\tiny recurrent$]

A classifier can always learn more causes, but more detailed studies are required in an urban road network to better characterize other causes of congestion. When a vehicle experiences excessive congestion, it exchanges via broadcast its vector \textit{C}. Each vehicle has its own decision module for the cooperative evaluation of the cause of congestion experienced. The decision module contains one of the Voting Procedure (VP) or Belief Functions (BF) methods described in the following sections.

\textit{A. Voting procedure}

The aim of this procedure is for vehicles to vote about the cause of congestion. Each vehicle has the probabilities vector presented above with one cause of congestion having the highest probability. They exchange their vector and collect others on the road segment via V2V communication. VP module computes the counts for each cause. The cause having the highest count is highlighted by this voting procedure as being the cause of congestion on the road segment.

The VP is an improvement of the  Back-Propagation algorithm presented in \cite{7837642}. In BP vehicles wait a certain duration and transfer messages only if they have total knowledge about the cause. We briefly describe the rules of BP for the sake of readability.

\begin{enumerate}[label={[\arabic*]}, noitemsep]
\item Vehicle \textit{i} continuously broadcasts and receives BEACON messages, computes current travel time, trajectory speed, demand and gap between vehicles;
\item If observed travel time is above threshold, Vehicle \textit{i} creates a feature vector and predicts with a classifier the cause of congestion;
\item Vehicle creates and propagates backwards a message called Event Request (RQ). RQ is retained locally on the segment for a minimum duration before propagating it to others and adjacent segments;
\item Once the duration is reached, Vehicle \textit{i} propagates Event Response (RP) message to adjacent road segments. 
\end{enumerate}

Nonetheless, if vehicles vote, some vehicles cannot quantify their ignorance on the presence or the absence of congestion. Belief functions avoid this problem because partial or total ignorance can be represented. 

\textit{B. Belief functions}

The aim of this method is to improve the level of knowledge. Classic probability theory is not a plausible model of how people make decisions based on weak information. The principle of indifference states that in the absence of information about some quantity \textit{X}, we should assign equal probability to any possible value of \textit{X}. The main objection against the use of classic probability theory as a model of epistemic uncertainty is its inability to represent ignorance. The theory of belief functions extends both the set-membership approach and classic probability theory \cite{4632299}. Let $\Omega$=\{$\omega$\textsubscript{1}, $\omega$\textsubscript{2}, ..., $\omega$\textsubscript{k}\} denotes a finite set containing all the possible answers to a given question \textit{Q} of interest; $\Omega$ being called the \textit{frame of discernment}. Information given by different sources regarding the answer to question \textit{Q} can be represented by a mass function, denoted by \textit{m}. It is defined from 2\textsuperscript{$\Omega$} (the set of all possible subsets of $\Omega$) to [0,1] such that the sum of all the masses is equal to 1:

\begin{equation}					
\sum_{A \subseteq \Omega  } m(A)=1
\end{equation}

A mass \textit{m(A)} represents the belief supporting \textit{A}, where \textit{A} is a subset of $\Omega$. It is the mass allocated to the hypothesis: the answer to question \textit{Q} belongs to the subset \textit{A} of $\Omega$. Each subset \textit{A} of $\Omega$ such that \textit{m(A)}> 0 is called a \textit{focal element} of \textit{m}. The theory allows the allocation of belief to subsets of $\Omega$ with no influence on the singletons, contrary to the probability theory \cite{5940448}. The mass \textit{m($\Omega$)} represents the degree of ignorance of the source which has provided the information \textit{m}. The mass on the empty set \textit{m($\emptyset$)} represents the conflict.

In case two vehicles express their beliefs over the frame, to combine the independent sets of probability mass assignments, \textit{m1$\cap$2} , quantified by \textit{m\textsubscript{1}} and \textit{m\textsubscript{2}} and expressed on $\Omega$, Dempster's rule of combination is the appropriate fusion operator. This rule derives common shared belief between multiple sources and ignores all the conflicting belief through a normalization factor. Specifically, the combination is calculated as follows:

\begin{equation}					
m1\cap2(A)= \frac{1}{1-K}\sum_{B \cap C = A } m_1(B)m_2(C), \forall A \subseteq \Omega
\end{equation}

where

\begin{equation}					
K = \sum_{B \cap C = \emptyset } m_1(B)m_2(C)
\end{equation}

\textit{K} is a measure of the amount of conflict between the two mass sets. With this combination, masses are transferred to focal elements intersections. The pignistic transformation \textit{BetP} transforms a normalized mass function \textit{m} into a probability measure as follows: 

\begin{equation}					
BetP(\{ \omega \})=\sum_{A \subseteq \Omega, \omega \in A} \frac{m(A)}{\mid A \mid (1-m(\emptyset))}, \forall \omega \in \Omega
\end{equation}

Applied to our problem, a mass function \textit{m} is held by each vehicle and is defined on the frame of discernment $\Omega$=\{Incident, Workzone, Weather, SpecialEvent, Recurrent\}. The classification model implemented on board of each vehicle returns the cause having the greatest probability, and the 2-item subset containing that singleton and its probability. We limit the strategy to 2-items because results showed very little improvement in the accuracy of prediction when more items are considered. In the next section, we propose to collect the 2-items subset for the sake of learning. 

\textit{C. Data association technique}

Once vehicles decide collectively via VP or BF about the cause of congestion, they can further apply some rules extracted by the data association technique presented in this section to improve their decision. We collect the vector of probabilities exchanged in many scenarios to build a dataset. We extract general association rules from the messages exchanged by analyzing the messages for frequent patterns in order to identify the relationships for rule generation \cite{anand2018association}. Let \textit{Ims} be the set of all possible items 
\textit{Ims} = \{ i\textsubscript{1}:Incident , i\textsubscript{2}:Workzone, i\textsubscript{3}:Weather, i\textsubscript{4}: Special Event, i\textsubscript{5}: Recurrent\}. Also, let \textit{t\textsubscript{i}} be a subset of items called itemset, \textit{T} is the set of all transactions T=\{ t\textsubscript{1}, t\textsubscript{2}, ... t\textsubscript{N}\} and \textit{N} being the total number of transactions in the dataset. Let support indicate how frequent items appear in the dataset i.e. number of transactions that contain a particular itemset \textit{X}.

\begin{equation}
\sigma(X)=|\{t\textsubscript{i}| X \subseteq t\textsubscript{i}, t\textsubscript{i} \in T\}| 
\end{equation}

Association rules are \textit{if/then} relationships that help uncover seemingly unrelated data in a relational database. There are two parts, the antecedent is the \textit{if}, found in the data and the consequent is the \textit{then}. If \textit{X} and \textit{Y} are disjoint sets \textit{X} $\cap$ \textit{Y} = $\emptyset$, for the rule  \textit{X}$\rightarrow$ \textit{Y}, we use support and confidence to identify relationships. 

\begin{itemize}

\item support of X$\rightarrow$ Y:

\begin{equation}
s(X \rightarrow Y)= \sigma(X \cup Y) / T  
\end{equation}

\item confidence of X$\rightarrow$ Y: number of times if/then statement have been found true. 

\begin{equation}
c(X \rightarrow Y)= \sigma(X \cup Y) / \sigma(X) 
\end{equation}

It is how frequently items in \textit{Y} appear in transactions that contain \textit{X}. It measures the reliability of the inference made by a rule. 

\end{itemize}

\begin{figure*}[!t]
 \begin{center}
 \includegraphics[width=0.9\linewidth,height=8cm]{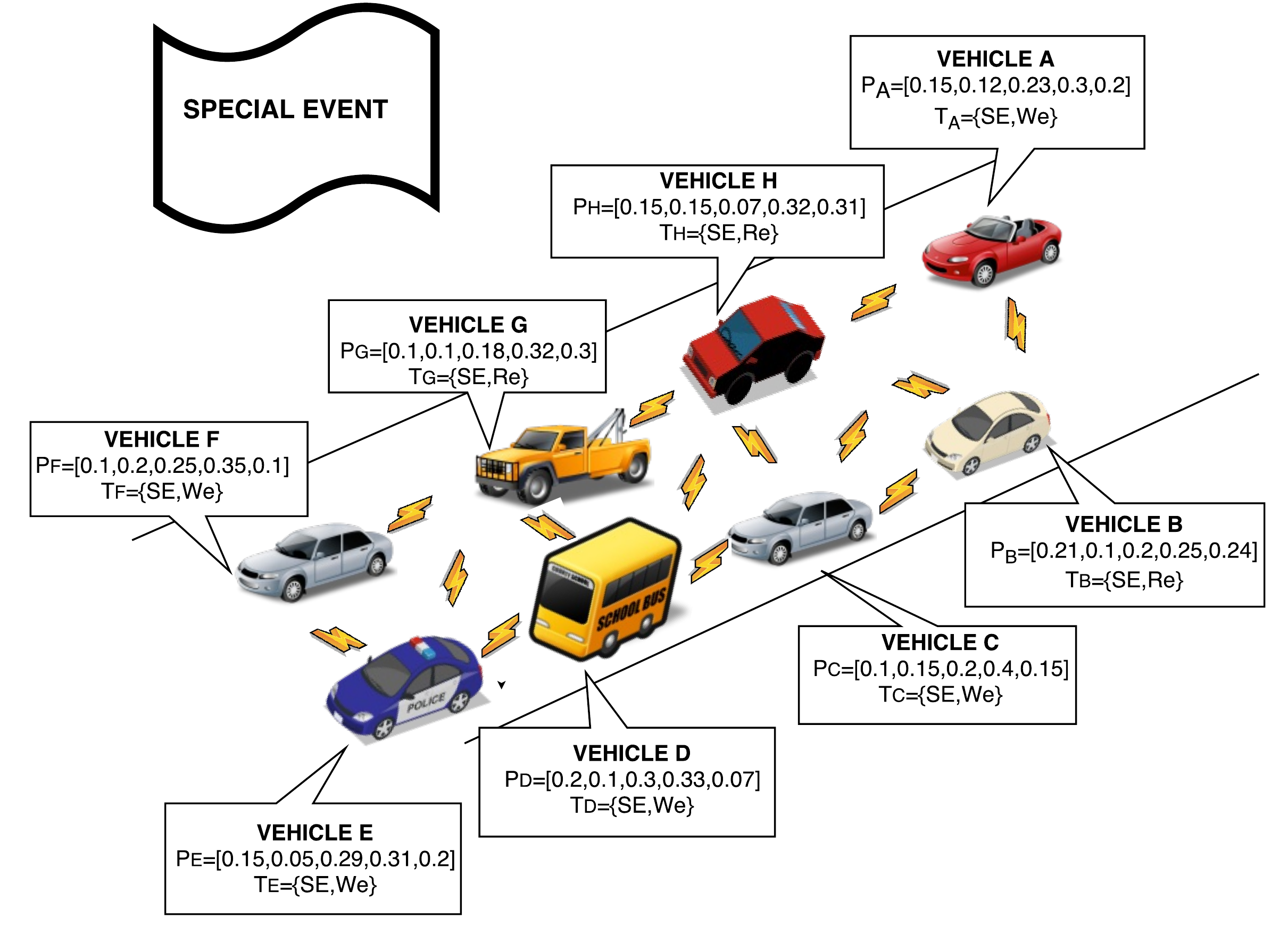}
 \caption{Transactions created by vehicles on a congested road segment due to a Special event}
 \label{fig:Transa_SE}
 \end{center}
\end{figure*}

The problem can be stated as follows: Given a dataset, find all the rules having support >= \textit{minsup} and confidence >= \textit{mincon},  \textit{minsup} and \textit{mincon} are thresholds derived from the dataset. If the probabilities vector [P$\tiny incident$, P$\tiny workzone$, P$\tiny weather$, P$\tiny specialevent$, P$\tiny recurrent$] of a vehicle is [0.15, 0.12, 0.23, 0.3, 0.2], the highest probability 0.3 corresponds to a special event (SE). The second probable cause as evaluated by the vehicle is a weather condition (We) with a probability of 0.23. The vehicle creates a transaction of categorical elements ordered as follows {SE, We}. In Fig.\ref{fig:Transa_SE}, we illustrate transactions created by many vehicles on a congested road segment due to a Special event. The problem can be decomposed into two major subtasks.

\underline{Frequent itemset generation}: We analyze data for frequent patterns by determining support count in the transactions for each candidate itemset in order to find all frequent itemsets that satisfy the minimum support threshold. Support-based pruning is done first to trim the exponential search space. We adapt the pseudocode for the frequent itemset generation part of the Apriori algorithm. Upon completion of this step, the set of all frequent 1-itemsets, F\textsubscript{1}, will be known. 

F\textsubscript{k}= \{i| i $\in$ Ims $\wedge$ $\sigma$(\{i\}) >= N*minsup\}

By replacing $\sigma$(\{i\})>=threshold with max($\sigma$(\{i\})), we find all maximum 1-itemsets as per the voting procedure described in the previous section.

Next, the algorithm will iteratively generate new candidate k-itemsets using the maximum (k-1) itemsets found in the previous iteration. Candidate 2-itemsets are generated using only the frequent 1-itemsets because the Apriori principle ensures that all supersets of the infrequent 1-itemsets must be infrequent:

F\textsubscript{k}= \{c| c $\in$ C\textsubscript{k} $\wedge$ max($\sigma$(\{c\})) \}

\underline{Rule generation}: The objective is to extract all the high-confidence rules from the frequent itemsets found in the previous step. Confidence-based pruning compares rules generated from the same frequent itemset to consider rules that have higher confidence. 

The implementation of the VP, BF and DAT in real conditions is presented in the following section.

\section{IMPLEMENTATION AND RESULTS}

The TAPAS Cologne scenario is considered to be one of the largest traffic simulation dataset. Using SUMO, we create extended scenarios mounted on top of the base scenario to model atypical traffic conditions. To simulate an Incident/Workzone, we stop on a lane some vehicles for a specific amount of time. In inclement weather, which lead to decreases in the vehicles' velocities and a more careful and defensive driver behaviour, we change the parameters of the car-following model in the simulator. To simulate a special event, we generate trips to a particular destination, with random departures and random routes. We use a Poisson process to generate random timings for trips. Table \ref{tableone} contains a description of the experiments in each scenario used for synthetic training set generation. In total, 10 experiments are investigated. Each experiment is conducted on 15 different road segments for the incident and the workzone scenarios and data of independent vehicles passing on the congested segments of each experiment in each scenario are evaluated. Experiments were conducted on 25 different road segments for the weather and the special event scenarios. 

\begin{table}[H]
\renewcommand{\arraystretch}{1}
\caption{Description of experiments}
\label{tableone}
\centering
\begin{tabular}{lllll} % Alignement  left dans toutes les colonnes
	\hline %Entte du tableau
	Scenario & Experiment & Description \\
	\hline %Contenu du tableau. Insertion ... & 3� colonne & 4� colonne - Incident at the beginning of a lane
	Incident 	& 1.1 &  Incident at beginning of lane \\
		    	& 1.2-1.3 &  At the middle-end of a lane\\
			& 1.4-1.5 &  For short-long duration\\
			& 1.6-1.7 &  Incident in-out Impact Region\\
	Workzone & 2.1 - 2.8  &  Same as incident experiments\\
	Weather 	& 3.1 &  Heavy weather condition\\
	Special Event & 4.1 - 4.4 &  Four different ingress flows \\
\end{tabular}
\end{table} %Fin du tableau

In SUMO, we generate urban mobility traces from the scenarios for usage in ns-2, the discrete-event network simulator. In the simulation of vehicular communications, we assume that vehicles are equipped with a Global Positioning System (GPS) device for positioning, a transceiver for communication using Dedicated Short-Range Communications (DSRC), and an enriched digital road map containing information about the map. We use broadcasting to pass information through the CVs. For communication among all cars, we assume standard signal range of the 802.11p protocol, which is 300 meters. BEACON messages are exchanged every 0.1 seconds.

The methods are validated by three indicators; estimation accuracy, detection time and percentage of false alarms. Accurate estimation of the root cause of congestion will enable road authorities to make more informed decisions about how to best reroute traffic across the multiple junction network as well as individuals adopting alternate routes. Also, lower detection time and false alarms will permit a rapid and exact reaction to resolve the traffic condition. The impact of delay in decision making is not minimal because traffic management operators can propose targeted mitigation strategies for traffic congestion based on the causes of bottlenecks with the aim of relieving traffic and increasing traffic flow resilience against cascading failures \cite{stergiopoulos2017interdependency}.

\subsection{Results}

\begin{figure*}[!t]
 \begin{center}
 \includegraphics[width=1\linewidth,height=7cm, keepaspectratio]{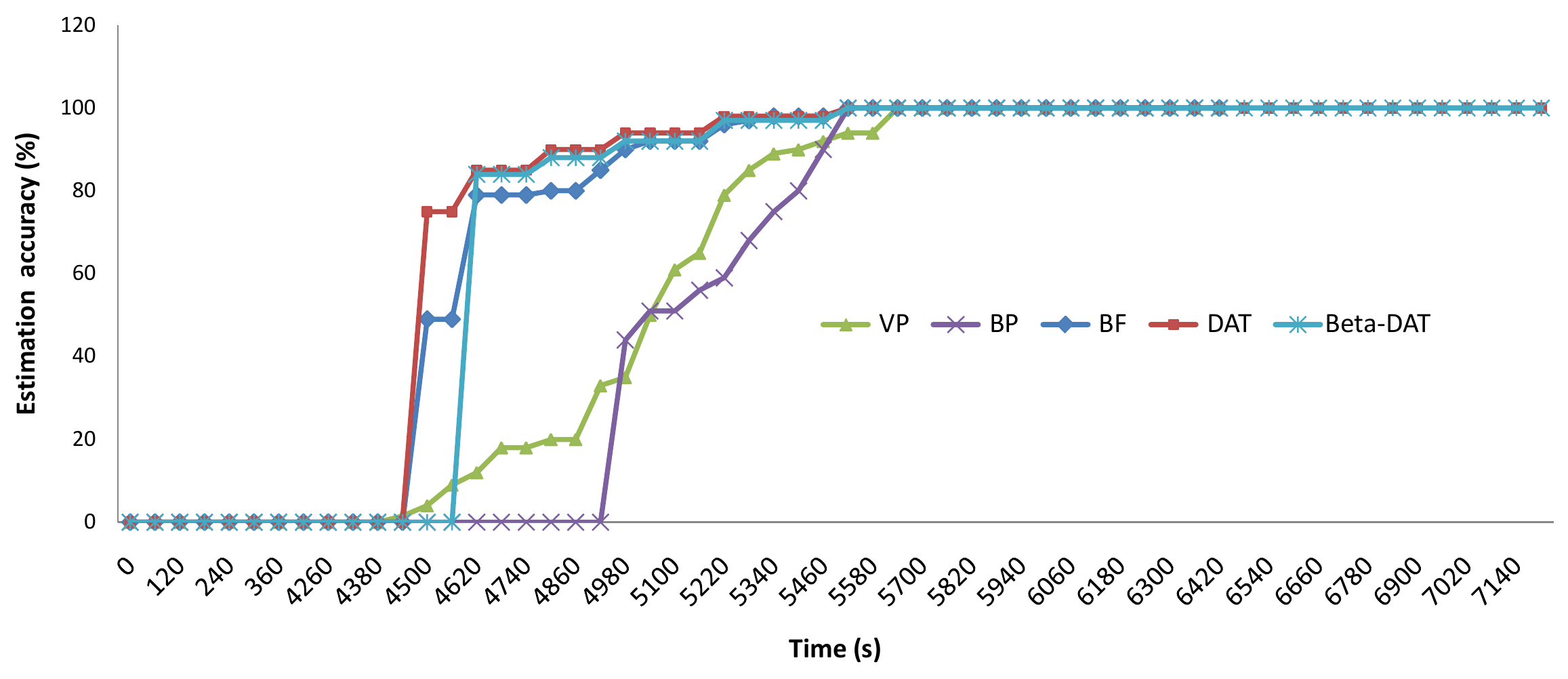}
 \caption{Estimation accuracy of our methods compared to BP in a weather condition }
 \label{fig:Performance_Estimation}
 \end{center}
\end{figure*}

Fig.\ref{fig:Performance_Estimation} compares the average percentage of vehicles accurately estimating the cause of congestion in different scenarios.  The best results the BP can do are after a certain time has elapsed because according to the algorithm, the first minutes following an excessive congestion not all vehicles on the segment are going to exchange messages regarding the event. At 4440s, vehicles start to vote via the VP. Only a few would have assessed that excessive congestion is present and will vote about the cause. Votes of all other vehicles on the segment are not counted because they did not have a probabilities vector for the estimation of the cause. At 4920s, when almost 39\% of the vehicles in VP voted about the cause of congestion, the BP algorithm activates and start signalling to vehicles that congestion is due to inclement weather. After performing many simulation runs from the experiments in Table \ref{tableone}, we found that the VP outperformed the BP method in every experiment in terms of percentage of vehicles accurately estimating the cause of congestion. On average, it did so by approximately 48\%. In terms of detection time, VP outperforms BP, it is reduced from 4920s to 4440s. This increase in performance is due to the fact that the VP method use CVs to elaborate a decision collectively on the cause of congestion. This shows that cooperation on a given geographical area to obtain deeper insight of traffic condition improves the accuracy of the estimation in contrast to BP that used CVs to disseminate the information and not for the sake of evaluation. One of the most difficult part of the simulation was finding road segments where an event would cause excessive congestion. On many segments, incidents might occur for a long duration but vehicles circumvent them. Our methods aim at detecting the cause of congestion and not detecting incidents or any other event. This analysis explains why it takes a relatively long time to achieve a desirable percentage of vehicles accurately estimating the cause of congestion in the scenario of Fig. 2.

With belief functions, partial knowledge or total ignorance can be represented. In Fig.\ref{fig:Performance_Estimation}, vehicles implementing BF exchange at the same time period messages assigning a mass for the singleton, another for a subset and an ignorance degree correspondingly. We report in Table \ref{tabletwo} results of vehicles S1-S22 exchanging messages M1-M22 at time t=4920s. Their mass functions have been combined using the conjunctive rule of combination resulting with a high confidence degree of 0.85 in the congestion caused by a weather event, and a low ignorance degree, with no conflict. Fig. \ref{fig:Performance_Estimation} shows that BF outperforms the VP and BP in terms of estimation accuracy.

\begin{table*}[!t]
\renewcommand{\arraystretch}{1}
\caption{Combination of mass functions m\textsubscript{i} from messages M\textsubscript{i},i $\in$ \{1,2,3,...,22\}}
\label{tabletwo}
\centering
\begin{tabular}{lllllllllllllll} % Alignement  left dans toutes les colonnes
	\hline %Entte du tableau
	Hyptothesis &  m\textsubscript{1}&  m\textsubscript{2} &  m\textsubscript{3} &  m\textsubscript{4}  & ... &  m\textsubscript{21} &  m\textsubscript{22} &  m\textsubscript{1$\cap$2 $\cap$3 $\cap$ ... $\cap$22} & BetP\\
	\hline %Contenu du tableau. Insertion ... & 3� colonne & 4� colonne - Incident at the beginning of a lane
	$\emptyset$ 	& 0 &  0 & 0 & 0 &  & 0 & 0 &0.652\\
	Incident & 0 &  0 & 0 & 0 &  &  0 & 0 &0.022&0.04\\
	Weather 	& 0.4 &  0.62 & 0.7 & 0.6 &  & 0 & 0.67 &0.1637&0.85\\
	Workzone 	& 0 &  0 & 0 & 0 &  &  0 & 0 &0\\
	Special Event & 0 &  0 & 0 & 0 &  &  0 & 0 &0\\
	Recurrent 	& 0 &  0 & 0 & 0 &  &  0.61 & 0 &0.1068&0.11\\
	\{Incident or Weather\} 	& 0 &  0 & 0.2 & 0.1 &  &  0 & 0.3 &0.0234\\
	\{Incident or Workzone\} 	& 0 &  0 & 0 & 0 &  & 0 & 0 &0\\
	\{Incident or SpecialEvent\} 	& 0 &  0 & 0 & 0 &  & 0 & 0 &0\\
	\{Incident or Recurrent\} 	& 0 &  0 & 0 & 0 &  & 0 & 0 &0\\
	\{Weather or Workzone\} 	& 0 &  0 & 0 & 0 &  &  0 & 0 &0\\
	\{Weather or SpecialEvent\} 	& 0 &  0 & 0 & 0 &  &  0 & 0 &0\\
	\{Weather or Recurrent\} 	& 0.3 &  0.3 & 0 & 0 &  &  0.34 & 0 &0.032\\
	\{Workzone or SpecialEvent\} 	& 0 &  0 & 0 & 0 &  & 0 & 0 &0\\
	\{Workzone or Recurrent\} 	& 0 &  0 & 0 & 0 &  & 0 & 0 &0\\
	\{SpecialEvent or Recurrent\} 	& 0 &  0 & 0 & 0 &  & 0 & 0 &0\\
	$\Omega$ 	& 0.3 &  0.08 & 0.1 & 0.3 &  & 0.05 & 0.03 &0.0000011\\
\end{tabular}
\end{table*} %Fin du tableau

Furthermore, we analyzed the messages exchanged by CVs for frequent patterns in order to identify relationships for rule generation. We use the default settings of the Apriori principle implemented in Weka \cite{Hall:2009:WDM:1656274.1656278}, a data mining software, to generate the rules. We then proceeded to the evaluation of the association rules. Association rule algorithms tend to produce a large set of rules. A rule is considered subjectively uninteresting unless it reveals unexpected information about the data. As domain experts, we extract the rules of interest because support and confidence measures are insufficient at filtering out uninteresting association rules. The rules extracted from the supervised dataset are: 

1. I$\rightarrow$ SE	
2. Wo$\rightarrow$ SE	
3. We,SE-$\rightarrow$ We	
4. Re,SE-$\rightarrow$ Re	
5. Re,We-$\rightarrow$ Re	 

Rules 1 to 5 are learned from the dataset. Rules 1 and 2 are extracted from the mining of the ordered items in each transaction, and we seize that when vehicles are experiencing congestion caused by an incident or a workzone, their second guess will never be anything other than a special event. Differently, rules 3-4-5 are extracted from the first item in each transaction and the label of the transaction. The consequent part of the rule is the label. Because each transaction is created by a vehicle in a specific scenario, the label of the transaction is the scenario. If the label is different from the first item in the transaction, it means that the vehicle wrongfully predicted the cause of congestion.  In Fig.\ref{fig:Performance_Estimation}, vehicles implementing DAT make use of the belief functions and the association rules uncovered to estimate the cause of congestion. On average the DAT improves estimation accuracy of 71\% compared to the BP method and 8\% compared to the BF. Detection time is decreased by 9.45\% informing of the congestion cause earlier. 

\begin{figure*}[!t]
 \begin{center}
 \includegraphics[width=0.9\linewidth,height=5cm]{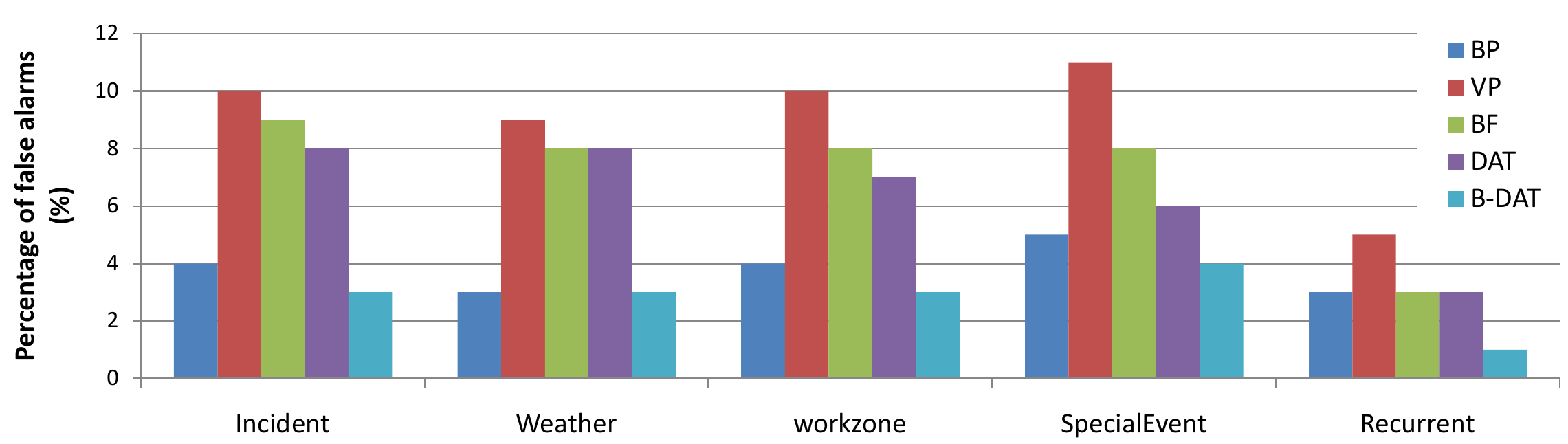}
 \caption{Percentage of false alarms of our methods compared to BP in different scenarios}
 \label{fig:PFA5}
 \end{center}
\end{figure*}

On the other hand, we compare the methods with the percentage of false alarms. In Fig. \ref{fig:Performance_Estimation}, false alarm is not included in the detection time. A false alarm is a vehicle initiating a VP or BP or DAT method and the simulation shows no excessive congestion, i.e. the situation captured by each vehicle is compared with the real simulation. Sending incorrect information may lead to the inaccurate assessment of the real traffic conditions. In case of a false alarms sent by the algorithms, the cooperation process and the evaluation layer we propose try to lower them. We monitored vehicles passing and report on average the percentage of false alarms triggered in Fig. \ref{fig:PFA5}. We found an average percentage error ranging from 3\% to 11\%. Although VP outperforms BP in terms of detection time and estimation accuracy, we notice that the voting procedure has the highest percentage of false alarms. 

Compared to the VP method, BF decreases the percentage of false alarms by approximately 1.8\% and DAT decreases it by approximately 1.33\%.Nonetheless, BP still outperforms DAT by approximately 3.25\% less false alarms in the incident, weather, work-zone and special-event scenarios. To investigate the situation further, we proceeded to a detailed monitoring of the incident scenario. We observed that false alarms happen earlier in the simulation and dissipate as congestion installs. We adapt the DAT method with an addition of a time factor, \textbeta. We call this method the \textbeta-DAT where we force the vehicles to wait for a certain time period, \textbeta, before cooperating for the evaluation of the cause of congestion. We found that the value of \textbeta has a direct impact on the percentage of false alarms and detection time and that it has no impact on the estimation accuracy. If the value of \textbeta is high, vehicles take as much time as BP to detect congestion and the percentage of false alarms tends to zero. With an upper bound for \textbeta being the duration of four consecutive link journey times as in \cite{anbarouglu2015non}, we conducted experiments and found that half of that duration is enough to attain the expected performance. 

In Fig.\ref{fig:Performance_Estimation}, \textbeta-DAT improved estimation accuracy by approximately 70\% compared to the BP method. Also, detection time of \textbeta-DAT is 7.09\% lower than that of the BP method, informing of the congestion cause earlier. It's 2.36\% higher than the DAT method, a slight increase of approximately 120 seconds. The consequence of adding a time factor to the DAT method on the detection time is insignificant compared to the benefit the duration added to the decrease of percentage of false alarms. In Fig. \ref{fig:PFA5} compared to the DAT method, \textbeta-DAT decreases the percentage of false alarms by approximately 3.6\%. Also, \textbeta-DAT outperforms BP by approximately 1.25\% less false alarms triggered by the network on the road segment. This shows that adapting the duration in combination with cooperation between CVs and knowledge on board of each vehicle improves overall performance for the accurate estimation of the cause of congestion. 

\subsection{Penetration rate of CVs}

\begin{figure*}[!t]
 \begin{center}
 \includegraphics[width=1\linewidth,height=9cm]{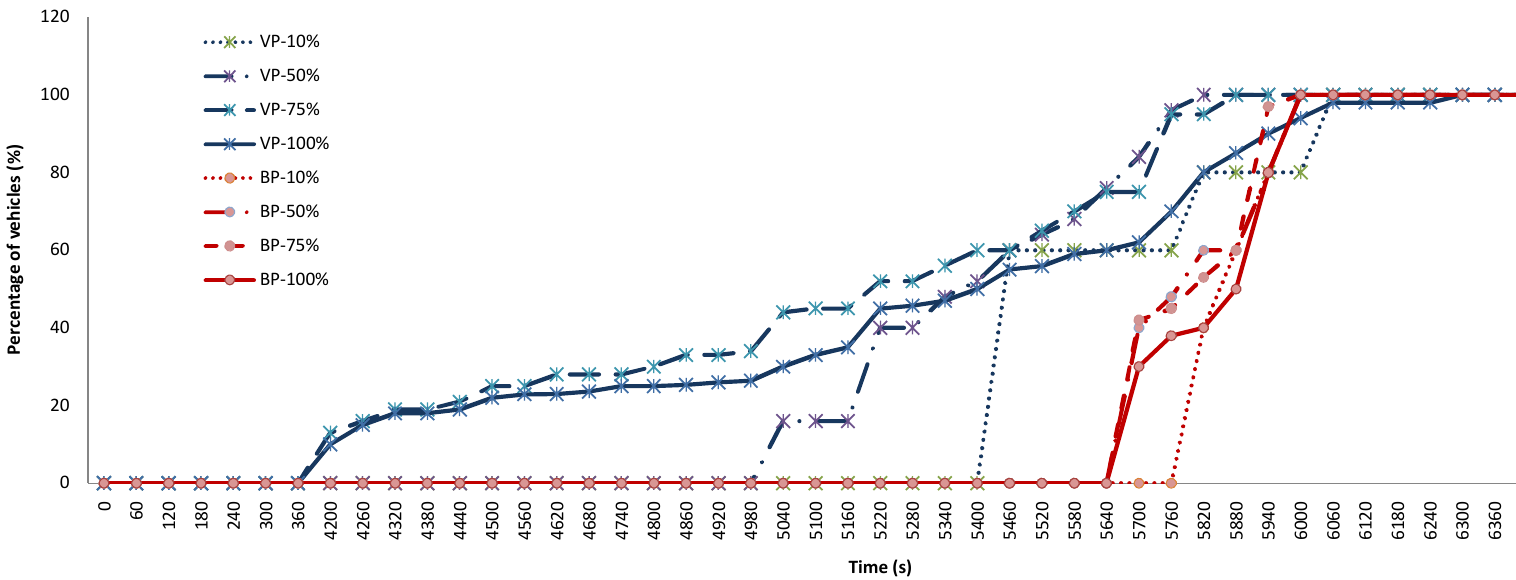}
 \caption{Impact of penetration rate on the performance of our method compared to BP in the incident scenario}
 \label{fig:PenetrationRate}
 \end{center}
\end{figure*}

The penetration rate in the first part of the study was 100\%. We now vary the number of participating vehicles and calculate the percentage of vehicles accurately estimating the cause of congestion as a function of time. We report the results in Fig. \ref{fig:PenetrationRate} for an incident scenario where the Y-axis shows the percentage of vehicles within the (10\%-50\%-75\% or 100\%) penetration rate of vehicles equipped with V2V.

A low penetration rate implies that fewer vehicles are equipped with transceivers and this leads to network fragmentation in VANET. In the BP method, vehicles have to wait to be in the communication range of each other while in our methods, fragmentation has a lower impact on the performance because they do not depend on network connectivity to resolve. We also notice that low penetration rates affect detection time more than estimation accuracy. At 10\% and 50\% penetration rate, vehicles detect that congestion is due to an incident only at 5460s and 5040s respectively. The question that arises in such a case is what percentage of penetration is enough to obtain the full benefits of the methods. We simulated many scenarios and on average, 63\% of penetration rate was acceptable in all scenarios except when the cause of congestion is a weather condition.

\begin{figure*}[!t]
 \includegraphics[width=1\linewidth,height=6cm, keepaspectratio]{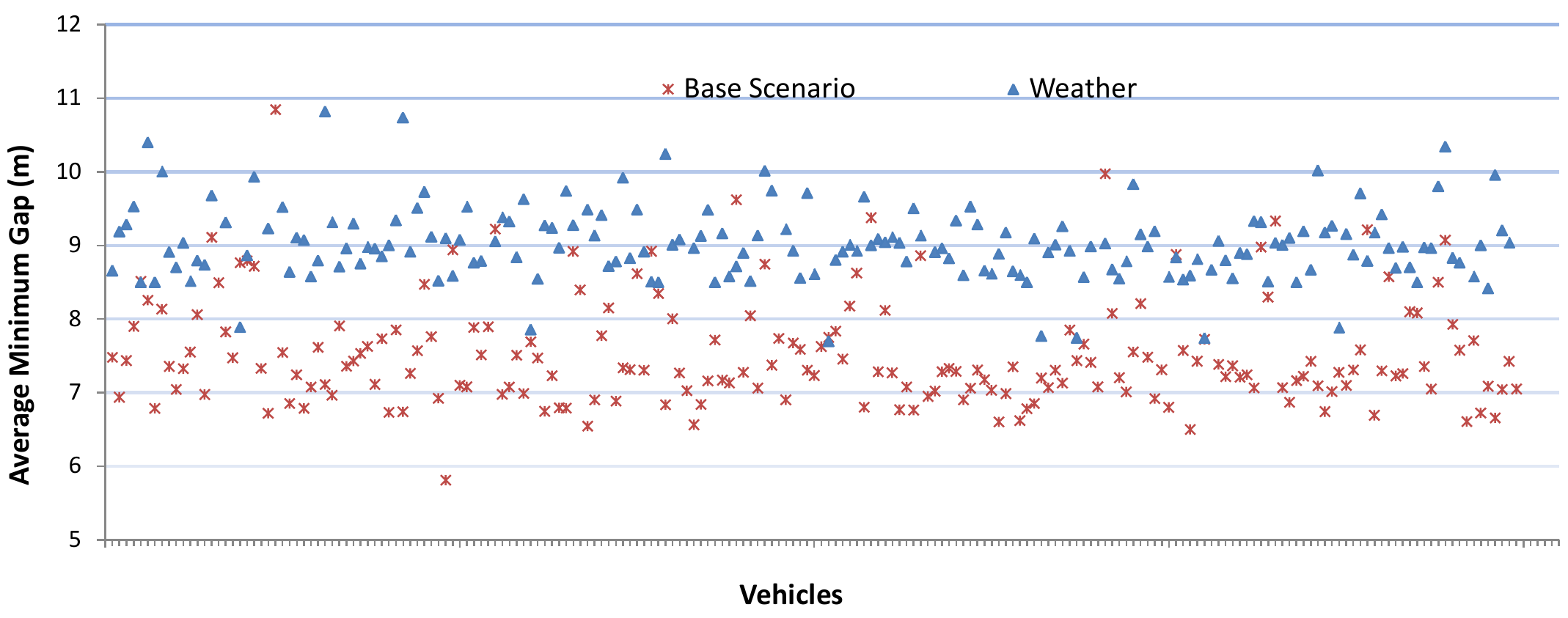}
\caption{ 85th percentile gap values in different scenarios}
\label{fig:GAP}
\end{figure*}

A very high percentage of CVs is required in order to achieve good performance in inclement weather. To study the situation further, we note in Fig. \ref{fig:GAP} the effect of weather on mobility by monitoring  the gap, empty space after the leading vehicle along a vehicle's trajectory. The desired following distance between two consecutive vehicles corresponds to the minimum safe gap attained by the vehicle. We notice from the \textit{85th} percentile gap measures collected by vehicles in the normal scenario and the weather scenario that in inclement weather, most vehicles augment their following distance. This leads to more network fragmentation and lower performance. Also, performance is very dependent on assumed parameters of the car-following model of the microscopic traffic simulator for the simulation of urban mobility in the inclement weather synthetic scenario. 

\section{CONCLUSION} 

Given that traffic involves multifaceted complex interactions, exploring the cause of congestion at a vehicle level is a partial and limited solution. In the event of a false alarm, spreading uncertainty among vehicles disrupts the proper network operation and presents a serious challenge in terms of security. We proposed methods using cooperation between CVs and added an evaluation process before data fusion. We also explored the collected data for learning purposes by building a dataset and extracting relationships and knowledge via a data mining technique. We tested and compared the methods using SUMO and ns-2, for the simulation of communication between CVs on real-world vehicular traces. Adapting the duration in combination with cooperation between CVs and knowledge on board of each vehicle enhances accuracy, decreases detection time and false alarms. Finally, our work has shown that our methods require approximately 63\% penetration rate to obtain the full benefits of CV communications. 

In a future work, one can proceed to a data network level investigation by adapting the transmission frequency and duration of messages exchanged specially in the presence of road traffic congestion. Network performance metrics such as packet delivery ratio, throughput and end-to-end delay can be examined. Since our methods are based on an infrastructure-less approach, it could be interesting to investigate how much of a lower penetration rate can be achieved with an infrastructure-based approach. Finally, to validate our scheme with the data on connected vehicles in the real-world conditions--in contrast to the synthetic dataset we used--could be an added value. 

% References section
\bibliographystyle{IEEEtran}
\bibliography{IEEEabrv,shortbibtexarticle}

% biography section
\begin{IEEEbiographynophoto}{Ranwa Al Mallah}
received the M.Sc. and Ph.D. degrees in Computer Science from Ecole Polytechnique de Montreal, Canada. Her current research interest includes Vehicular networks, and Traffic Efficiency and Safety Applications.
\end{IEEEbiographynophoto}

\begin{IEEEbiographynophoto}{Alejandro Quintero}
received the diploma of advanced studies, and the Ph.D. degree in computer engineering, respectively, from the INPG – Grenoble and Joseph Fourier University, Grenoble, France. He is currently a full professor at the Department of Computer Engineering of  Polytechnique de Montreal, Canada. 
\end{IEEEbiographynophoto}

\begin{IEEEbiographynophoto}{Bilal Farooq}
received his Ph.D. degree from University of Toronto, Canada in 2011, respectively. His current research goal is to develop multidisciplinary and highly-intelligent solutions for sustainable planning, design, and operations of urban infrastructure systems.
\end{IEEEbiographynophoto}

\end{document}